\documentclass[a4paper,fleqn,usenatbib]{mnras}

\usepackage{newtxtext,newtxmath}
\usepackage[T1]{fontenc}
\usepackage{ae,aecompl}

\usepackage{graphicx}	% Including figure files
\usepackage{amsmath}	% Advanced maths commands
\usepackage{amssymb}	% Extra maths symbols
\usepackage{units}
\title[Short title, max. 45 characters]{On the high energy cut-off of accreting sources: is GR relevant?}

\author[F. Tamborra et al.]{
Francesco Tamborra,$^1$\thanks{E-mail: francesco.tamborra@asu.cas.cz}
Iossif Papadakis,$^{2,3}$
Michal Dov\v ciak,$^1$, 
and Ji\v ri Svoboda$^1$ 
\\
$^{1}$ Astronomical Institute of the Czech Academy of Sciences, Prague, Czech Republic \\
$^2$ Department of Physics and Institute of Theoretical and Computational Physics, University of Crete, 71003, Heraklion, Greece \\
$^3$ IESL, Foundation of Research and Technology, 71110 Heraklion, Greece
}

\date{Accepted XXX. Received YYY; in original form ZZZ}

\pubyear{2017}

\begin{document}
\label{firstpage}
\pagerange{\pageref{firstpage}--\pageref{lastpage}}
\maketitle

% Abstract of the paper
\begin{abstract}
The hard X--ray emission observed in accreting compact sources is believed to be produced by inverse Compton scattering of soft photons arising from the accretion disc by energetic electrons thermally distributed above the disc, the so-called X-ray corona. Many independent observations suggest that such
coronae should be compact and located very close to the black hole. In this case general relativistic (GR) effects should play an important role to the continuum X-ray emission from these sources, and in particular in the observed high energy cut-off, which is a measure of the intrinsic temperature of the corona. Our results show that the energy shift between the observed and intrinsic high energy cut-off due to GR effects can be as large as 2 - 8 times, depending on the geometry and size of the corona as well as its inclination. We provide estimates of this energy shift in the case of a lamp-post and a flat, rotating corona, around a Kerr and a Schwartzschild black hole, for various inclinations, and coronal sizes. These values could be useful to correct the observed high energy cut-off and/or coronal temperatures, either in the case of individual or large sample of objects. 
\end{abstract}

% Select between one and six entries from the list of approved keywords.
% Don't make up new ones.

%%%%%%%%%%%%%%%%%%%%%%%%%%%%%%%%%%%%%%%%%%%%%%%%%%

%%%%%%%%%%%%%%%%% BODY OF PAPER %%%%%%%%%%%%%%%%%%

\section{Introduction}
Up-scattering of low-energy photons by Inverse Compton effect on a hot gas of 
electrons (i.e. Comptonization) is a common mechanism for high-energetic 
plasmas, especially for accreting sources such as Black Hole Binaries (BHBs) and Active 
Galactic Nuclei (AGN). It is widely believed to be 
the main source of the hard X-ray emission we observe in radio quiet sources (i.e. AGN without a jet): soft photons produced by the accretion disc are 
Comptonized by a medium of hot electrons (usually referred to as the 
``X--ray corona") whose geometry and physical parameters are mostly unknown. 
If the electrons are thermally distributed in the corona with a certain energy, 
$kT_e$, the resulting X-ray spectrum produced by Comptonization, usually 
referred to as the primary component, will be a power-law with a cut-off 
proportional to the thermal energy of the corona depending on the optical depth 
of the corona, $\tau$. For some specific geometry and optical thickness of the 
corona, this proportionality has been found to be $E_{cut} \sim 2-3$ $kT_e$ (e.g. \cite{POP2001}).

The primary component irradiates the inner accretion disc and produces
a reflection spectrum whose main features are the fluorescent Iron 
K$\alpha$ at 6.4--6.7 keV (depending on the ionization of the disc) and 
a 'Compton hump' due to reflection of $\gtrsim 10$ keV photons from the disc.
Since reflection may originate from disc parts which are close to the black 
hole (BH), general relativistic (GR) effects (most prominently the gravitational 
redshift) may affect the observed spectrum. The GR effects on the reflection 
X-ray spectral components have been extensively used for spin and corona 
size measurements (for redshifted iron line e.g. \cite{Fabian2002}, \cite{Miniutti2007}, and for redshifted Compton hump e.g. \cite{Svoboda2015}).
Most of these measurements have suggested the presence of highly spinning black 
holes and compact coronae in AGN.  X-ray spectral-timing and reverberation 
analysis of AGN spectra support this conclusion and in many cases the position 
of the corona is thought to be less than $3-10 r_g$ above the black hole 
(e.g. \cite{Kara2016},  \cite{Emmanoulopoulos2014}, \cite{DeMarco2011}).

In the last years, thanks to the large broadband observations obtained mainly 
by \textit{NuStar}, spectra up to hundreds of keV have been collected for 
many AGN and BHBs. Some of them show a clear high energy cut-off in 
the power-law, allowing to estimate the thermal energy of the corona, $kT_e$ (for an exhaustive collection of results consult \cite{Fabian2015} and \cite{Lubinski2016}).  If the X--ray source lies close to the BH, GR effects such as gravitational redshift, relativistic beaming (if the corona rotates) and light bending will become important and may affect the estimation of thermal energy of the corona derived from the observed high-energy cut-off.  GR effects to the X--ray primary component have already been studied in the case of point-like sources located on the axis of disc rotation by  \cite{Niedzwiecki2016}, \cite{Fabian2015} and \cite{Fabian2017}.

Point-like and extended disc-like coronae orbiting or rotating above the disc (e.g. \cite{Wilkins2012} and  \cite{Wilkins2016}) have been extensively studied in the last few years and GR corrections are usually taken into account for such geometries in order to properly calculate emissivity profiles from the disc reflecting the X-ray primary emission coming from such coronae as well as to properly model the expected X-ray reverberation signal. However, in many works involving spectral analysis only, GR corrections on primary component are neglected. 

The main goal of this work is to provide correction factors for the observed energy cut-offs in the case of X--ray spectra emitted by  a disc-like corona, i.e.  a flat, extended corona over the surface of the accretion disc. The spectrum emitted is affected mainly by gravitational redshift and Doppler shifts. We compute and provide such factors for a large number of coronal radii, inclinations, constant and radially decreasing emissivity profiles, for a maximally rotating and a Schwarzschild black hole. Depending on the inclination and the size of the corona, the intrinsic energy cut-offs can be significantly larger, but also {\it smaller}, than the observed ones. They can also be significantly larger than the respective correction factors in the case of point-like sources. Our results can be used in the study of the intrinsic energy cut-off (and hence coronal temperature and intrinsic thermal energy) distribution in AGN by simply apply the corrections we furnish to the observed energy cut-off derived by usual spectroscopical analysis. 
Comparison of the resulting intrinsic distributions may offer additional clues regarding the geometry of the X-ray sources in these objects.

\section{Relativistic effects to the continuum spectrum}
We assumed two compact, simple geometries for the X--ray emitting region. 
The first is the so-called lamp-post geometry where the source is static and 
located on the BH spin axis, at a height $h$ above the disc plane (\cite{Matt1991}). 
The second is a geometrically thin corona extended 
from the radius of the innermost stable circular orbit (ISCO) to an outer radius, $r_{out}$,
in the equatorial plane just above the accretion disc and rotating with 
Keplerian velocity around the BH (hereafter the 
``disc corona''). Both height and radii are measured in units of the 
gravitational radius, $r_g=GM/c^2$. While both these geometries are highly 
simplified, one can imagine the 
geometry of the real X--ray compact source to be in between these two extreme 
configurations (with $h$ and $r_{out}$ being representative 
of the source's height and extension, respectively).

To study the GR effects on the X--ray continuum spectrum, we assumed that the X-ray source emits isotropically (in the rest-frame) with an intrinsic spectrum of the form:

\begin{equation}
\label{intr-pl}
F(E)=NE^{-\Gamma}\exp(\dfrac{-E}{E^{i}_{cut}}).
\end{equation}

Then, we computed the spectrum that a distant observer would detect, by taking 
into account all relativistic effects for both geometries (gravitational and Doppler energy shifts and light bending (\cite{Karas2006}). In both cases  
the spectrum will be shifted in energy and its normalisation will change while the shape of the power-law spectrum, i.e. the power-law index, $\Gamma$, will remain the same. We are interested in particular in the energy shift of $E_{cut}$, since this observable is used to measure the thermal energy of the corona. We therefore define the $g$-factor\footnote{Note that the g-factor in GR is usually defined conversely but we find this definition more useful in our case. If g-factor is known, we can multiply the observed cut-off energy by this number to get the intrinsic cut-off energy.} as the ratio of the intrinsic, $E^{i}_{cut}$, to the observed cut-off energy, $E^{o}_{cut}$, e.g. 
\begin{equation}
\label{g-fact}
g = E^{i}_{cut}/E^{o}_{cut}.
\end{equation}

In the case of the lamp-post corona $g$ is independent of inclination (as only gravitational redshift contributes ) and therefore it can be calculated using the simple relation,

\begin{equation}
\label{g_LP}
g = 
    \sqrt{\dfrac{h^2+a^2}{h^2-2h+a^2}},
\end{equation}
where $h$ is the height of the corona above the BH, and $a$ is the non-dimensional  spin parameter. This is a well known formula, and has already been used in the past, for example, by \cite{Fabian2015} and \cite{Fabian2017} to correct the observed cut-off energies in their sample.

In the case of a disc corona, we expect the $g-$factor to depend on the radius of the corona, its emissivity profile, as well as the inclination. However, due to the fact that the primary power-law cut-off emission from different parts of the disc corona will be shifted by different $g-$factors due to the fact that both the gravitational redshift as well as the Doppler shift (corona rotates) vary across the corona, the overall $g-$factor, as defined by eq.~(\ref{g-fact}), cannot be estimated analytically. We give below an example of how we can estimate $g-$factor in this case.

Let us consider a disc corona which extends from the horizon of a maximally rotating black hole (spin $a = 1$) to 2.5 $r_g$, with uniform emissivity and an intrinsic spectrum (in its rest frame) defined by eq.~(\ref{intr-pl}), for $\Gamma=2$ and $E^{i}_{cut}=100$ keV (green solid line in Fig.~\ref{spectrum}).  The red solid line in the same figure shows the integrated spectrum, as it would be detected by a distant observer with an inclination\footnote{We define inclination as the angle, $\theta$, between the line of sight and the BH spin axis} of $\theta=30 ^{\circ}$. Both spectra were artificially renormalized to the value of 1 at 1 keV, to better see the energy shift and the shape of the observed spectrum. We then fitted the observed spectrum (over the $1-100$ keV band) with a model of the form,

\begin{equation}
\label{obs-pl}
F_{obs}(E) = NE^{-\Gamma}\exp\left(g\,\dfrac{-E}{E^{i}_{cut}}\right).
\end{equation}
\noindent
{The normalisation constant, $N$, power-law index, $\Gamma$, and the energy shift of cut-off, $g$, were free parameters during the fitting process. The blue dashed line in Fig.~\ref{spectrum} shows the best-fit line. The best-fit $g-$factor is 2.08. In other words, for this disc corona, the observer will detect an energy cut-off which will be {\it half} of $E^i_{cut}$. Strictly speaking, the shape of the observed spectrum (red line) is not identical to a power-law cut-off model as defined by eq.~(\ref{obs-pl}). The drop due to the cut-off energy in the observed spectrum is slightly broadened (relativistically smeared), compared to an exponential cut-off. Nevertheless, for all practical purposes, the exponential cut-off model (blue dashed line) describes the shape of the observed spectrum very well. We therefore accept the best-fit $g-$value as representative of the $g-$factor for this disc corona. 

\begin{figure}
\includegraphics[width=88.mm,clip=]{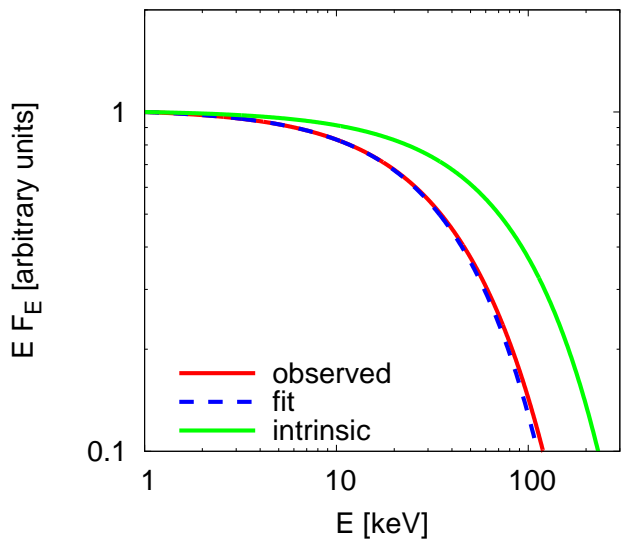}
\caption{\label{spectrum} The intrinsic spectrum emitted from a disc corona 
extending from the horizon of a maximally rotating black hole (spin = 1) to 2.5 
$r_g$ (red line), and the spectrum detected by a distant observer  at  
$30 ^{\circ}$  from the spin-axis (green line). The dashed-blue line indicates 
the best-fit of the function defined by eq.~(\ref{obs-pl}) to the observed 
spectrum.}
\end{figure}

\section{The observed high-energy cut-off}

In order to determine the $g-$factor in the case of disc coronae, we followed the procedure described in the previous section for a large number of $r_{out}$ and inclinations.  We considered an extreme Kerr ($a=1$) and a Schwarzschild BH ($a=0$), with an inner disc radius of 1 and 6 $r_g$, respectively. We also considered the case of a disc corona with a constant emissivity and a radially reduced emissivity, as $r^{-3}$. In all cases we assumed that the intrinsic spectrum of the corona is described by eq.~(\ref{intr-pl}). We estimated the spectrum that would be observed at infinity (the ``model-observed spectrum"), taking into account all the GR effects, and we renormalized its value to 1 at 1 keV. We fitted it with the model defined by eq.~(\ref{obs-pl}), and we accept the best-fit $g-$value ($g_{fit}$) as the $g-$factor for the disc corona, with any given $r_{out}, \theta$, and emissivity profile.
 
During the fits, we always let $N$ and  $\Gamma$ as free parameters but, in any case,  $N_{fit}$ and $\Gamma_{fit}$ would always be almost identical to 1 and to the intrinsic slope, $\Gamma^i$. We fitted the model-observed spectra in the energy range 1--100 keV, which is broadly similar to the energy range that people would use in practise, with combined $XMM-$Newton and $NuSTAR$ data. Increasing (up to 1 MeV) or decreasing (down to 0.2 keV) the limits of the fitting energy band does not affect the resulting $g_{fit}$. However, the best-fit $g-$value changes slightly if we move the lower energy limit from 1 to 3 keV. The difference is less than 8\% in the case of very compact coronae ($r_{out}< 1.5 \, r_g$), and only for a few inclinations (in most cases the difference is less than $\sim 2-3$\%). 

Since the model-observed spectrum is not entirely identical to the intrinsic spectrum, as we discussed above, we tried various $\Gamma^i$ and $E^i_{cut}$ values for all the $r_{out}$ and $\theta$ combinations we considered. In this way we can investigate how robust the $g_{fit}$ values are to different spectral shapes. We tried the values of $E^i_{cut}=30, 100, 300$ and 1 MeV, and we found that $g_{fit}$ does not vary substantially, as long as $E^i_{cut}>100$ keV. For smaller intrinsic cut-off values, we observed a difference by no more than $\sim 8$\% (in the case of the most compact disk coronae). In the case of different intrinsic spectral values, we found that $g_{fit}$ increases by $14$\% (at most), from $\Gamma^i=3$ to $\Gamma^i=1$, when $r_{out}<1.5$.

Table \ref{table:1} lists our results in the case of a disc corona, with $\Gamma^i=2$ and $E^i_{cut}=100$ keV. Rows list the $g-$factor for different inclinations ($1^{\circ}<\theta<85^{\circ}, \Delta\theta=1^{\circ}$) and columns list the $g-$factor for various $r_{out}$ (from 1 or 6 $r_g$ up to to 20 $r_g$). Both the constant and $r^{-3}$ emissivity results are listed. The lamp-post results are given in the last row for different corona heights (equal to the $r_{out}$ listed on the top row). Their estimation is simply based on eq.~(\ref{g_LP}), but we list them for completeness. Given the differences we found for various $\Gamma^i$, we provide similar Tables for $\Gamma^i=1, 1.5, 2, 2.5, 3, 3.5$ (and  $E_{cut}^{i}=100$ keV), online.
 
\begin{table}
\caption{$g-$factors for disc coronae with $\Gamma^i=2, E^i_{cut}=100$ keV, and constant/radial power-law ($r^{-3}$) emissivity, for different observer inclination (rows), and corona outer radius (columns). Lamp-post (LP) $g-$factors are listed in the last row, for different heights (equal to $r_{out}$, as listed in the the first row). The full table is available on line. }
\label{table:1}     
\centering         
\begin{tabular}{l | c | c | c | c | c}
\hline
\hline
$r_{out}$   & 1.52 & 2.50 & 5.05  & 10.2  & 20.0  \\
$\theta$ &  &  &  &  &  \\
\hline
\vspace{2mm}
 5$^\circ$&$\nicefrac{6.00}{6.21}$&$\nicefrac{2.66}{2.88}$&$\nicefrac{1.61}{1.84}$&$\nicefrac{1.27}{1.52}$&$\nicefrac{1.14}{1.40}$\\
\vspace{2mm}
30$^\circ$&$\nicefrac{4.26}{4.41}$&$\nicefrac{2.08}{2.24}$&$\nicefrac{1.40}{1.58}$&$\nicefrac{1.18}{1.38}$&$\nicefrac{1.09}{1.31}$\\
\vspace{2mm}
60$^\circ$&$\nicefrac{2.10}{2.18}$&$\nicefrac{1.24}{1.33}$&$\nicefrac{1.04}{1.14}$&$\nicefrac{1.00}{1.09}$&$\nicefrac{1.00}{1.08}$\\
\vspace{2mm}
80$^\circ$&$\nicefrac{1.15}{1.19}$&$\nicefrac{0.88}{0.93}$&$\nicefrac{0.89}{0.91}$&$\nicefrac{0.93}{0.91}$&$\nicefrac{0.96}{0.92}$\\
\hline
LP & 3.50 & 1.80 & 1.27 & 1.11 & 1.05 \\
\hline
\hline
\end{tabular}
\end{table}

\begin{figure}
\includegraphics[width=88.mm,clip=]{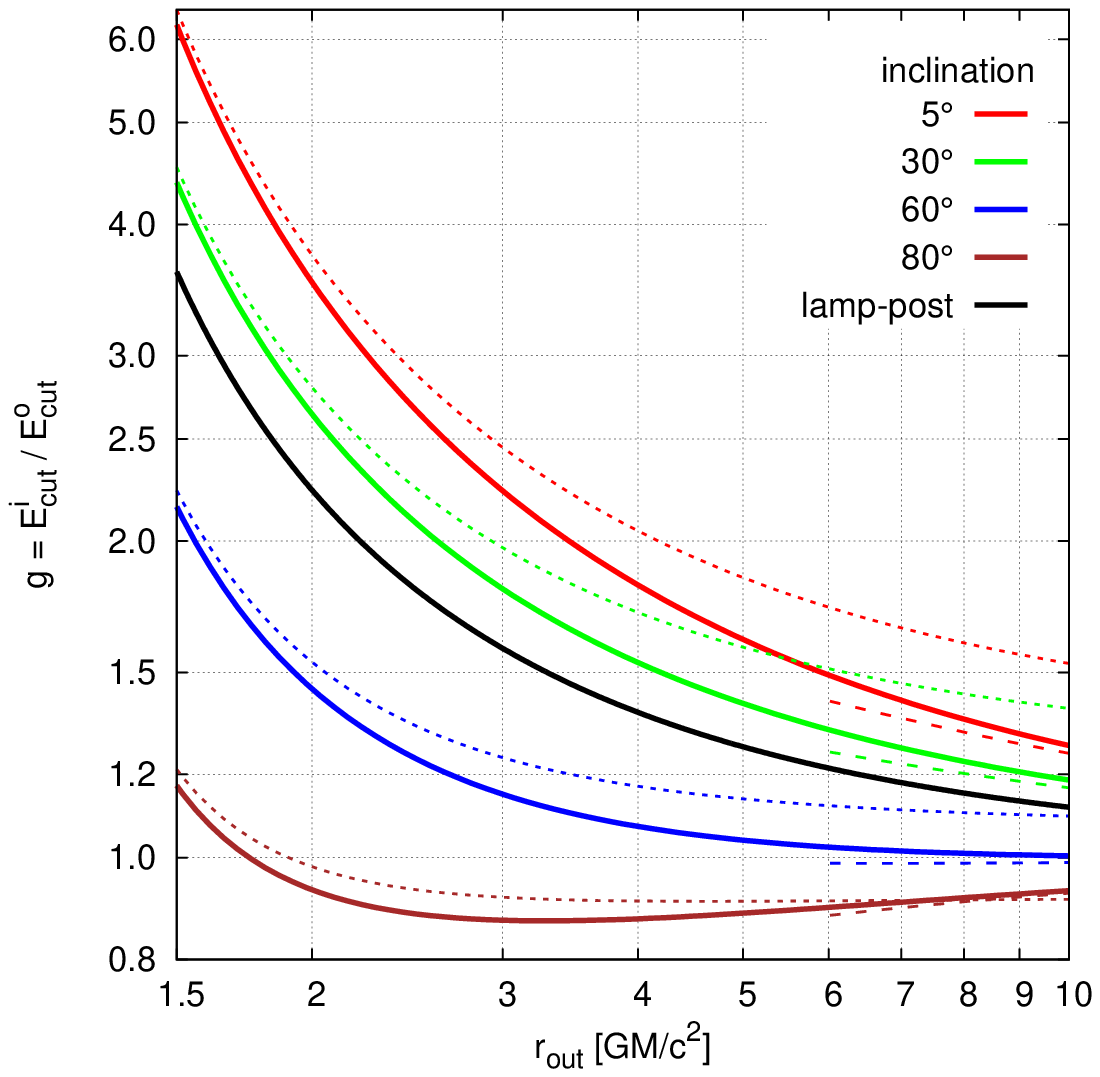}
\includegraphics[width=88.mm,clip=]{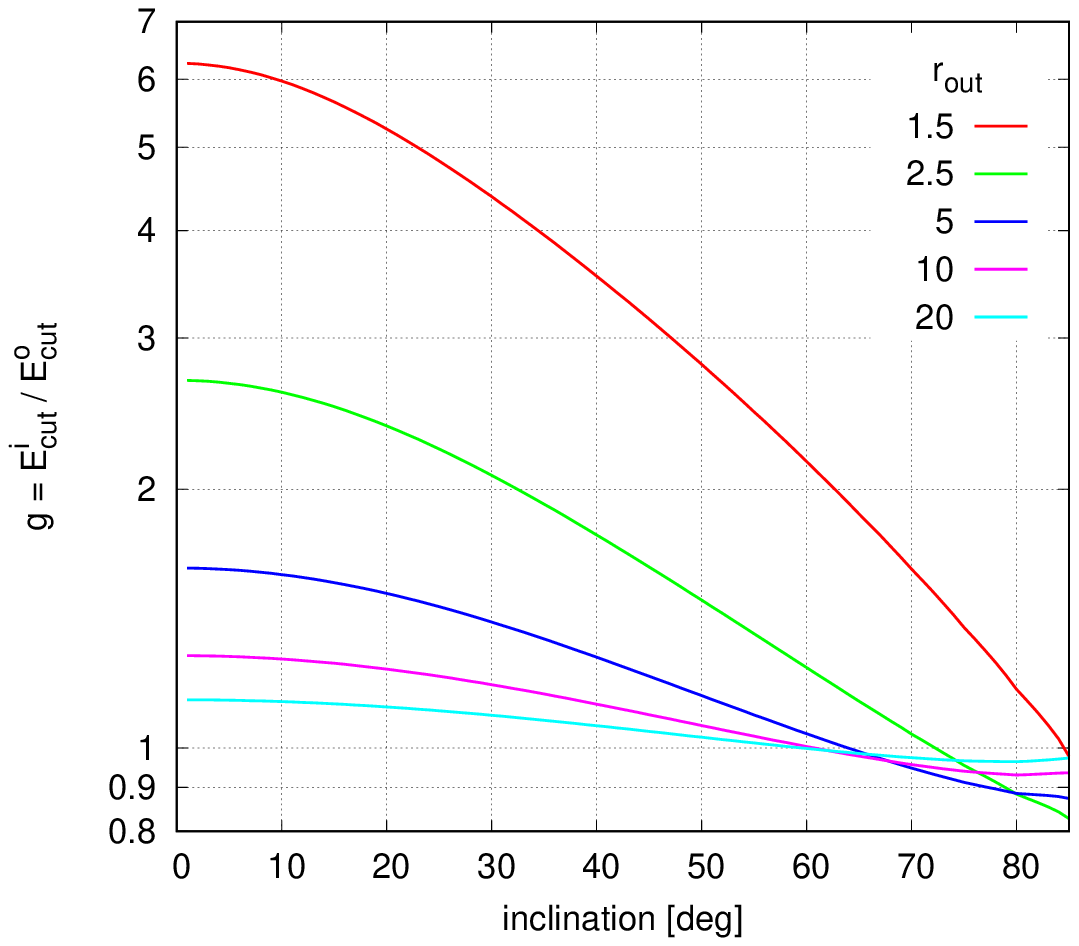}

\caption{\label{g}{\it Top panel:} The $g-$factor plotted as function of $r_{out}$, for various inclinations, in the case of a disc corona with $\Gamma^i=2, E^i_{cut}=100$ keV, constant emissivity for extreme Kerr and Schwarzschild BH (solid and dashed lines, respectively) and in the case of a disc corona with a $r^{-3}$ emissivity profile (dotted lines). The black solid line shows $g-$factor in the case of lamp--post geometry (in this case, the $r_{out}$ values on $x-$axis correspond to $h$). 
{\it Bottom panel:} Plot of $g-$factor as a function of the inclination for 
several $r_{out}$ in the case of disc coronae with constant emissivity.}
\end{figure}

Fig.~\ref{g} shows a summary of our results (listed in Table \ref{table:1}). In the lamp-post case (solid 
black line in the top panel), the $g-$factor does not depend on the inclination, 
and it is always larger than one, since gravitational redshift is the main GR 
effect that affects the spectrum. As a result, the energy of all spectral 
features should be smaller in the observer's frame. The effects to the spectrum 
become noticeable when the source is very close to the BH. When $h<3 \, r_g$, 
the observed cut-off energy could be 40\% smaller than the intrinsic value. 
It could be less than half, when $h<2 \, r_g$. 

In the case of disc corona, the $g-$factor is determined mainly by gravitational 
redshift and Doppler shift. The results for disc coronae with a $r^{-3}$ emissivity profile and with uniform emissivity are plotted with the dotted and solid lines, respectively. For the same $r_{out}$, the $g-$factors are always larger in the former case. This is due to the fact that  the effective radius of the $r^{-3}$ coronae is much smaller than $r_{out}$. The difference increases with increasing $r_{out}$. 

The effect of Doppler shift dominates at large 
inclinations ($\theta>60^{\circ}$; blue and brown lines in top panel of Fig.\,\ref{g}). In fact, at very high inclinations, the observed high energy cut-off could be up to $\sim 20 \%$ {\it larger} than 
$E_{cut}^i$ (peaking at $r_{out} \sim 3 r_g$). On the other hand, the $g-$factor 
could be as high as $2-6$ in the case of X--ray sources which are seen face on 
($\theta<30^{\circ}$) and are close to the BH ($r_{out}\lesssim 2.5 r_g$; bottom panel in Fig.\,\ref{g}). 

In the case of Schwarzschild BHs (dashed lines in the top panel of 
Fig.~\ref{g}), the contribution to large $g-$factor  from radii smaller than $6 \, r_g$ is missing. Therefore, the overall $g-$factor is smaller than  the $g-$factor in the case of disc coronae around maximally rotating BHs, with the same radius. The effect decreases with increasing outer radius,  since the relative contribution of the inner disc decreases with increasing coronal size (mainly in the case of the constant emissivity disc coronae). 

\section{Discussion}

In this paper we studied in detail how GR effects affect the observed cut-off energy (and hence temperature) measurements in AGN  the case of disc-like coronae. Many studies of extended X--ray coronae in AGN have been published the last few years. For example, \cite{Wilkins2012} proposed a disc-like corona in 1H 0707-495, which is located as low as $2 r_g$ above the plane of the accretion disc and extends outwards from the rotation axis to around $30-35 r_g$, based  on the comparison between observed emissivity profiles to those computed theoretically for different locations and geometries of the X--ray source, and on the study of the time lags spectrum of the source. This X--ray corona geometry is almost identical to the geometry we study in this work: a flat corona, located above the disc, with an inner radius of $r_{ISCO}$ (for a maximally rotating and a Schwarzschild BH) and an outer radius of $r_{out}$. 

We computed $g-$factors for coronae with constant emissivity and with an emissivity that decreases with radius as $r^{-3}$, for a large number of inclinations ($1^{\rm o}<\theta<85^{\rm o}$) and corona outer radius ($1.1 r_g \le r_{out} \le 20 r_g$). Our estimates are valid for any intrinsic cut-off energy (as long as $E^i_{cut}\ge 30$ keV). We used a simple phenomenological model for the continuum spectrum (namely a simple power-law with high energy cut-off; \textit {cutoffpl} in XSPEC terminology). This is not identical to more physical Comptonization models such as \textit{comptt} or \textit{compps} (e.g. \cite{Matt2015}, \cite{Marinucci2014}). We chose \textit{cutoffpl} because it was easy and fast to compute our results. Nevertheless, the differences between this model and other Comptonization models is much smaller than the difference between \textit{cutoffpl} with $\Gamma^i=3$ and $\Gamma^i=1$ that we considered in our study. Therefore, the ratio of the intrinsic to the observed roll-over energy at high energies should always be given by the $g-$factors listed in Table \ref{table:1}, irrespective of the model that is used to determine $E^o_{cut}$ (or $kT_e^{o}$). 

The $g-$fit values depend, slightly, on $\Gamma^i$. We observe a moderate difference of $\sim 14$\% in the $g-$fit values in the case when $\Gamma^i=1$ and 3 (and $r_{out}$ is smaller than 1.5 $r_g$). These are rather extreme situations, and the $g-$factors differ by a much smaller amount in all other cases. Nevertheless, we provide $g-$factors for a large number of intrinsic slopes, although the corrections we list in Table\ref{table:1} should be adequate in most practical situations. 

In the case of lamp-post geometry, GR corrections to $E^i_{cut}$ are easily applicable, and have already been discussed by \cite{Fabian2015} and \cite{Fabian2017}. We provide $g-$factors in this case as well, for completeness. Our results show that the $g-$factor for low inclination disc coronae ($\theta<30^{\circ}$) is larger than the lamp-post $g-$factor (for the same $h$ and $r_{out}$), even in the case of constant emissivity (solid green and red lines in the top panel of Fig.\,\ref{g}). 
This is due to the fact that the contribution to the corona emission from radii below $r_{out}$ is quite substantial. While the Doppler shift is low at these face on systems, the strong gravitational redshift effects at small radii can produce stronger cut-off energy shifts. On the other hand, $g-$factors in disc coronae are smaller than the $g-$factors for point-like sources in high inclination systems, when Doppler effects dominate. 

In general, we find that the GR effects to the X--ray continuum spectrum can be quite significant. If the X--ray corona in AGN is extended over the disc and $r_{out}$ is small, $E^i_{cut}$ (and hence the temperature) has been systematically underestimated. For example, in the case of disk coronae with $\theta<40^{{\rm o}}$ (i.e. Type I AGN) and $r_{out}<2.5 r_g$, $E^i_{cut}$ will be at least 2 times larger than $E^o_{cut}$. The difference could be even larger than 6 in the case of ultra compact coronae in face-on objects. The difference between $E^i_{cut}$ and $E^o_{cut}$  is less than 1.5 when $\theta > 60^{\rm o}$ (this corresponds to Type II AGN). In fact, at inclinations larger than 70 degrees, the intrinsic cut-off energy could be {\it smaller} then $E^o_{cut}$, by up to 20\%. 

So far, cut-off energies are estimated by fitting observed spectra without applying GR corrections to the continuum models. As we argued above, irrespective of the continuum model used, the $g-$factors we provide can be multiplied with the observed $E^o_{cut}$ in order to estimate the intrinsic cut-off energies (or temperatures) to estimate $E^i_{cut}$. Given the differences between the $g-$factors in the case of point-like and disc-like, extended coronae, it would be useful to apply both corrections, when comparing estimated temperatures with theoretical predictions (i.e. with the pair production limit). In both cases, it is necessary to assume the coronal size (either $h$ or $r_{out}$) and $\theta$. In fact, comparison of the resulting $E^i_{cut}$ with maximum allowed temperatures in X--ray corona, can put constrains to the geometry of the source (i.e. $r_{out}$ and/or $h$, depending on the assumed geometry), in addition to any other constrains that may have been estimated by fitting X--ray reprocessing features in the energy spectra, or by timing studies. 

In addition to a source by source basis, our results could also be used in other ways to investigate the X--ray source in AGN. For example, cut-off energies have been measured in many AGN in the last 10--20 years or so. The results show a distribution of values  between $\sim 40-200$ keV (even higher in some cases). Our results show that, even if the intrinsic cut-off energy is the same in all AGN, there should be a considerable scatter in the observed energies, if the objects are observed  at various inclinations and/or the outer radius of the corona is not the same in all sources (see bottom panel in Fig. 2).  This possibility could explain some, or even all of the scatter in the observed cut-off energies in AGN.

To investigate this issue, we considered the results of \cite{Lubinski2016}. They analysed the hard X-ray spectra of 28 bright Seyfert galaxies observed with {\it INTEGRAL}, together with data from {\it XMM-Newton, Suzaku} and {\it RXTE}, and they determined the  mean coronal temperature ($kT_e$) in these objects. The solid black line in the upper panel of Fig.\,\ref{resplot} shows the distribution of the observed temperatures, $kT_e^{o}$ (Table 3 of \cite{Lubinski2016}). 

\begin{figure}
\includegraphics[width=88.mm,clip=]{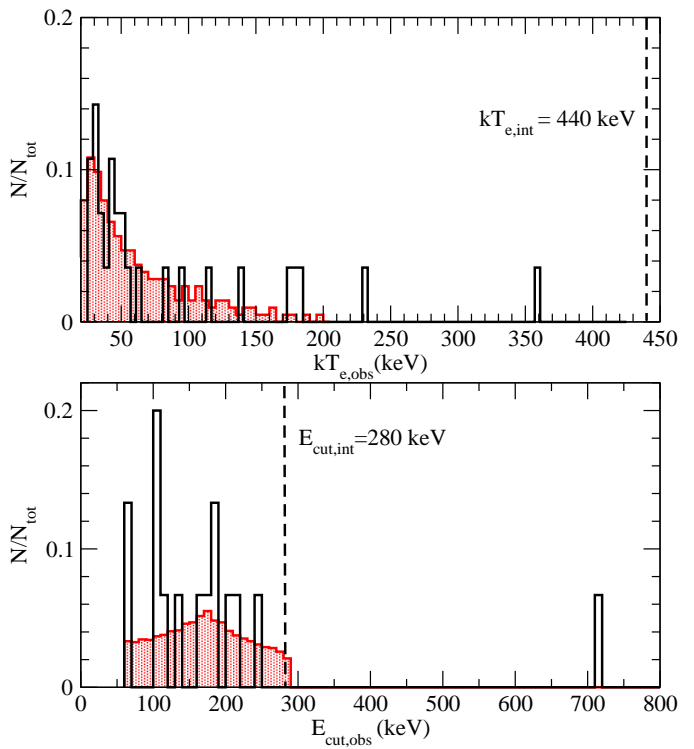}
\caption{\label{resplot} The distribution of the observed $kT_e$ and $E_{cut}$ in AGN (black histograms in the upper and lower panels, respectively) from the
\protect\cite{Lubinski2016} and \protect\cite{Fabian2015}
$NuStar$ sample.  The red histogram in both panels shows the predicted $kT_e$ and $E_{cut}$ values that we would observe in the case when the intrinsic temperature/cut-off energy (indicated by the vertical dashed lines in the two panels) and coronal size is the same in all AGN (see text for details)
}
\end{figure}

In order to investigate whether the distribution of the observed temperatures could be consistent with the hypothesis of a single intrinsic temperature, we considered a range of intrinsic temperatures, $kT_e^{i}$, from 200 to 500 keV (with a step of 5 keV), and we used the $g-$factors listed in Table \ref{table:1} (in the disc corona case, with uniform emissivity), to predict the distribution of the observed temperatures ($kT_{e,mod}^{o}$) for every $kT_e^i$ value. We let $\theta_{min}$, $\theta_{max}$ and $r_{out}$ to be free variables in the estimation of the $kT_{e,mod}^o$ distribution, and we used the K-S test to compare each model distribution with the distribution of $kT_e^{o}$. We found that the observed temperature distribution is fully consistent ($p_{null}=0.8$) with the distribution of temperatures that we would observe  if $kT_e^i=440$ keV,  the corona is very compact in all objects ($r_{out} \le1.2 r_g$), and their inclination is uniformly distributed between $13^{\circ}<\theta<85^{\circ}$ (red histogram in the upper panel of Fig.\,\ref{resplot}). The average luminosity of the objects in the \cite{Lubinski2016} sample is $\sim 0.08$, in Eddington luminosity units (from their Table 5). Therefore, if the corona is less than 1.2 $r_g$, the average X--ray compactness in these objects, $l$,  should be larger than 1500. However,  the point $(\Theta\sim440/512=0.86 \, l=1500)$ lies above all pair lines in the $(\Theta,l)$ plane (see, e.g., Fig.\,2 in \cite{Fabian2015}).

We therefore conclude that the AGN in the \cite{Lubinski2016} sample cannot have the same intrinsic cut-off energy, as in this case the X--ray corona would be too small, and its compactness unphysically  too high.

We performed the same analysis using the observed cut-off energies of the {\it NuStar} sample of AGN in \cite{Fabian2015}. This sample is smaller than the \cite{Lubinski2016} sample, but {\it NuStar} data can provide accurate measurements of $E_{cut}$. The solid black line in the lower panel of 
Fig.\,\ref{resplot} shows the distribution of $E_{cut}^i$ for this sample. The red histogram in the same panel indicates the distribution of the cut-off energies that we would observed if $E_{cut}^i=280$ keV and $r_{out}= \, 5r_g$ in all AGN, and the inclination of the objects is uniformly distributed between $17^{\circ}<\theta<77^{\circ}$. The model and the observed $E^i_{cut}$ distributions are fully consistent ($p_{null}=0.89$\footnote{Due to the small number of objects, and the weakness of the K-S test to account for differences in the wings of the distribution, the very large cut-off energy of NGC~5506 at $\sim 700$ keV does not affect the goodness of fit between the two distributions.}). Due to the higher lower limit in $\theta$, and the larger coronal size, the model distribution in this case is flatter than the model distribution plotted in the upper panel of Fig.\,\ref{resplot}. Assuming a factor of 2 to convert from the cut-off energy to the coronal temperature (this is the case if the optical depth is $\lesssim 1$, \cite{POP2001}) then $\Theta=0.27$ for these AGN. If the average X--ray luminosity of the objects in this sample is $\sim 0.08$ (same as before), then for a $5\, r_g$ corona, the compactness will be $l\sim 400$. 

We therefore conclude that the intrinsic cut-off energy  can be $\sim 280$ keV in all AGN in the \cite{Fabian2015} sample.
The radius of the X--ray corona should be $\sim 5 r_g $ (in agreement with the results from recent spectral/timing studies in many of them), and the resulting $(\Theta,l)$ values will locate these AGN close to the pair line for a hemispherical corona (Fig. \,2 in \cite{Fabian2015}).

The analysis above is meant to be an example of how our results could be used in the study of the properties of AGN coronae.  The results are not very restrictive at the moment, but we should be able to perform a much more detailed analysis in the near future, as the number of sources with well measured high energy cut-offs should increase, thanks to {\it NuStar}, but also {\it Swift/BAT},  observations.

\section*{Acknowledgements}
We acknowledge financial support from the European
Union Seventh Framework Programme (FP7/2007-2013) under grant agreement n.312789.
MD would like to thank for the support from the project 17-02430S funded by Grant Agency of the Czech Republic.
JS acknowledges financial support from grants LTAUSA17095 and Inter-Inform LTI17.

%%%%%%%%%%%%%%%%%%%%%%%%%%%%%%%%%%%%%%%%%%%%%%%%%%

%%%%%%%%%%%%%%%%%%%% REFERENCES %%%%%%%%%%%%%%%%%%

% The best way to enter references is to use BibTeX:

\bibliographystyle{mnras}
\bibliography{newbib} % if your bibtex file is called example.bib

% Alternatively you could enter them by hand, like this:
% This method is tedious and prone to error if you have lots of references
%\begin{thebibliography}{99}
%\bibitem[\protect\citeauthoryear{Author}{2012}]{Author2012}
%Author A.~N., 2013, Journal of Improbable Astronomy, 1, 1
%\bibitem[\protect\citeauthoryear{Others}{2013}]{Others2013}
%Others S., 2012, Journal of Interesting Stuff, 17, 198
%\end{thebibliography}

%%%%%%%%%%%%%%%%%%%%%%%%%%%%%%%%%%%%%%%%%%%%%%%%%%

%%%%%%%%%%%%%%%%% APPENDICES %%%%%%%%%%%%%%%%%%%%%

%%%%%%%%%%%%%%%%%%%%%%%%%%%%%%%%%%%%%%%%%%%%%%%%%%

% Don't change these lines
\bsp	% typesetting comment
\label{lastpage}
\end{document}